\documentclass[a4paper,fleqn]{cas-dc}

\usepackage[numbers]{natbib}

\def\tsc#1{\csdef{#1}{\textsc{\lowercase{#1}}\xspace}}
\tsc{WGM}
\tsc{QE}

\begin{document}
\let\WriteBookmarks\relax
\def\floatpagepagefraction{1}
\def\textpagefraction{.001}
\let\printorcid\relax 

\shorttitle{Design and Implementation of a Lightweight Object Detection System for Resource-Constrained Edge Environments}    

\shortauthors{Zhengbao Yang et al.}

\title[mode = title]{Design and Implementation of a Lightweight Object Detection System for Resource-Constrained Edge Environments
}

\author[1]{Jiyue Jiang}

\author[1]{Mingtong Chen}

\author[1]{Zhengbao Yang}
\cormark[1] 
\ead{zbyang@hk.ust} 
\ead[URL]{https://yanglab.hkust.edu.hk/}

\address[1]{The Hong Kong University of Science and Technology
Hong Kong, SAR 999077, China}

\cortext[1]{Corresponding author} 

\begin{abstract}
This project aims to develop a system to run the object detection model under low power consumption conditions. The detection scene is set as an outdoor traveling scene, and the detection categories include people and vehicles. In this system, users’ data does not need to be uploaded to the cloud, which is suitable for use in environments with portable needs and strict requirements for data privacy. The MCU device used in this system is STM32H7, which has better performance among low-power devices. The YOLOv5 system is selected to train the object detection model. To overcome the resource limitation of the embedded devices, this project uses several model compression techniques such as pruned, quantization, and distillation, which could improve the performance and efficiency of the detection model. Through these processes, the model's computation and the quantity of model parameters could be reduced, in order to run computer vision models on micro-controller devices for the development of embedded vision applications.

\end{abstract}




\maketitle

\section{Introduction}

In recent years, artificial intelligence (AI) has made significant progress in various fields such as computer vision, natural language processing, and speech recognition. It has been widely applied in areas including intelligent security, autonomous driving, and smart home systems. As the performance of AI models continues to improve, their structures have become increasingly complex, with a growing number of parameters and higher computational demands. Consequently, the deployment of these models relies heavily on high-performance hardware. However, such devices are often bulky, costly, and high energy consumption. Furthermore, users’ data needs to upload to cloud in these systems, making them unsuitable for applications that require energy-efficient, potable and privacy solutions. 

To address these limitations, edge computing is becoming one of the core solutions. Deploying AI models on edge devices can significantly reduce the latency and bandwidth consumption associated with transmitting data to the cloud, thereby enhancing both real-time performance and intelligence levels. Moreover, users’ data doesn’t need to upload to cloud, this method could efficiently protect customer’s privacy. However, most edge devices have very limited of computational ability, traditional object detection models such as YOLO (You Only Look Once) and SSD (Single Shot Multibox Detector), which excellent in accuracy and performance, require high computational resources, making them difficult to deploy on resource-constrained edge devices.The emergence of TinyML (Tiny Machine Learning) technology addresses this challenge by compressing and optimizing machine learning models, enabling them to run efficiently on low-power, low-storage embedded devices, opening up new possibilities for smart edge devices. These devices are highly suitable for use in environments where portability is important and data privacy is critical.

This project aims to carry out object detection on low-power, portable devices. Using deep learning model on SCM (Single Chip Microcomputer) to detect two categories: people and vehicles. Based on YOLOv5 pre-trained model, using a dedicated dataset with two target detection categories to train the detection model, then using pruning, quantization, techniques to compress model. Finally, deployed and executed the optimized model on STM32H743 device.

\section{HARDWARE DEVICE and MODEL SYSTEM SELECTION}

In order to achieve a better detection effec, the H7 series MCU device with higher performance in the STM32 series was chosen for this project. The STM32H7 series is ST Microelectronics’ family of Cortex‑M7‑core microcontrollers designed for high‑performance embedded applications. Building on the existing STM32 ecosystem, it delivers higher clock speeds, larger on‑chip memory capacities, and a richer set of peripherals to meet the stringent compute and real‑time requirements of industrial control, robotics, audio/video processing,  AI inference, and similar use cases. The specific device is STM32H743XIH6, with Arm Cortex-M7 chip, 2M FLASH storage, 1M RAM storage, and the maximum frequency is up to 480 MHz. In the this project, no other hardware devices such as gas pedals are used.

The nominal RAM storage space of the STM32H743XIH6 device is 1M and the FLASH storage space is 2M, but the actual available RAM space could be used by users is about 640KB, because a certain amount of space is needed to run the inference and the application of the HAL library requires a certain amount of space for storage and operation. The inference operation needs to be executed on the RAM, and, in addition, the data from the camera data acquisition also needs to be temporarily stored in RAM.The weight parameters of the AI model are stored in FLASH.

In this project, the camera module and display module used in the selection of the more commonly used models. The camera module model is OV5640, with 5 mega-pixel, using the DCMI interface communication; The display module is 2.0-inch TFT LCD screen, using the SPI interface communication.

\begin{figure}[h]
	\centering
		\includegraphics[scale=0.8]{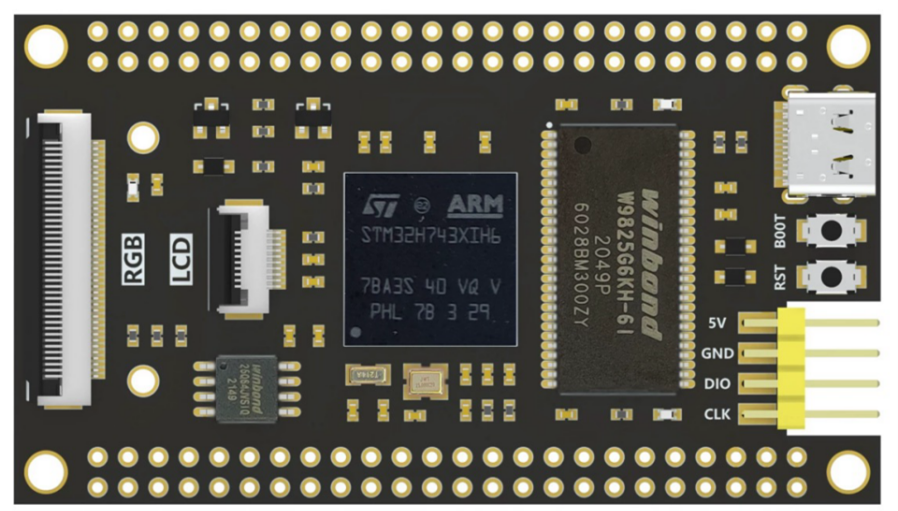}
	  \caption{Hardware device: STM32H743XIH6}\label{FIG:1}
\end{figure}

\begin{figure}[h]
	\centering
		\includegraphics[scale=1]{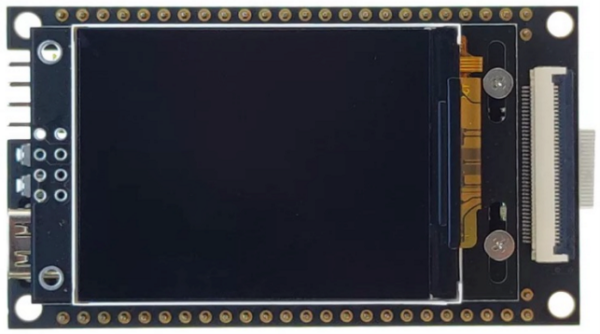}
	  \caption{Hardware device:TFT LCD screen}\label{FIG:2}
\end{figure}

\begin{figure}[h]
	\centering
		\includegraphics[scale=1]{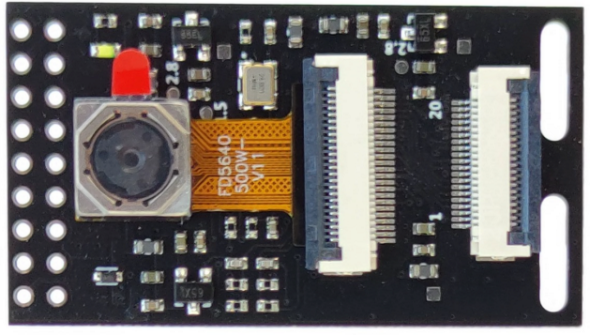}
	  \caption{Hardware device: OV5640}\label{FIG:3}
\end{figure}

There are many object detection models in the field of computer vision. The current deep learning-based object detection field can be broadly classified into two types, two-stage detection algorithm and one stage detection algorithm.
Two stage detection algorithm: The basic principle of this type of algorithm is to first form a number of columns of candidate regions as the material to be detected, and then input these regions to be detected into the Convolutional Neural Network for further detection or classification\cite{1},Representative models such as Faster R-CNN, Mask R-CNN, etc.;One stage detection algorithms: The advantage of this type of algorithm is that there is no need to generate a candidate box, can directly picture the kind of objects that need to be detected into the regression of the problem to look at, can be only one step can be completed on the detection of the target[1]. Using this method could obtain a short time to performed detection task, however the accuracy is not good as two-stage detection algorithms. Representative models include YOLO series, SSD, RetinaNet and so on. Researchers mainly use two-stage detection at first, represented by the R-CNN series, through the generation of candidate regions to achieve high-precision detection, but this detection method is computationally complex, and the detection speed is slow. This problem is solved by one-stage detection algorithms. The number of parameters for one-stage detection is smaller, and the real-time response ability is stronger. With these advantages, single-stage detection gradually replaces two-stage detection, and becomes the main model category of target detection at this stage.The first one-stage detection model is YOLOv1 in 2016, the recent version of YOLO model is YOLOv11, released by 2024. The YOLO family of models has become the dominant model for applications in one-stage detection because of its end-to-end design and lightweight improvements to achieve real-time performance while maintaining high accuracy.
Among YOLO models, YOLOv5 achieves a good balance between detection speed, detection accuracy, and model complexity.\cite{2}

YOLOv5 is an efficient target detection algorithm, especially in real-time target detection tasks. This model handles large, medium, and small objects with three different scales of detection heads; the detection heads consist of three key tasks: bounding box regression, category prediction, and confidence prediction; each head uses three anchors on a pixel-by-pixel basis to help the algorithm predict the object boundaries. Depending on the number of channels and parameters in the model structure, the YOLOv5 pre-training model has several subdivided versions, from small to large, namely YOLOv5n, YOLOv5s, YOLOv5m, YOLOv5l, and YOLOv5x, which can be adapted to different types of tasks and hardware platforms.\cite{3} YOLOv5n has the least parameters, the model parameters of YOLOv5n is 1.9M, not exceed 2.0M, is suitable for use in STM32H743 hardware device. Therefore, YOLOv5n model is selected for this project for target detection and deployment of the application.
 
All of the YOLOv5 models share the same model structure. The model structure is mainly composed of backbone, neck and head, as shown in Fig. 4.The backbone is use for feature extraction, including Conv, C3, SPP/SPPF modules. Conv module combines convolution, batch normalization (BN), and SiLU activation, used for downsampling and adjusting channel dimensions; C3 module consists of multiple residual (Bottleneck) blocks with skip connections, merging shallow and deep features to prevent gradient vanishing; SPP/SPPF used to rapidly fuses multi‑scale features via serial max‑pooling (5×5 kernels), increasing the receptive field.\cite{4}
The neck part is use for multi-scale feature fusion, the structure mainly consists of CBS, upsample, and C3 modules. The feature fusion network in YOLOv5 mainly adopts a structure that combines FPN and PAN. FPN primarily extracts feature maps from the feature extraction network, progressing from lower layers to higher layers, followed by upsampling and feature fusion. Based on FPN, PAN adds an additional bottom-up path, which allows positional information from lower layers to be transferred more quickly to higher layers.\cite{5} This part could improve detection performance for small, medium and large objects.
The head in YOLOv5 serves as the detection and output result component of the algorithm. Its core function is to perform localization and classification planning on the predicted bounding boxes. Generally, the localization process of the predicted boxes is treated as a regression problem, which relies on a specific loss function.\cite{2}

\begin{figure}[h]
	\centering
		\includegraphics[scale=0.65]{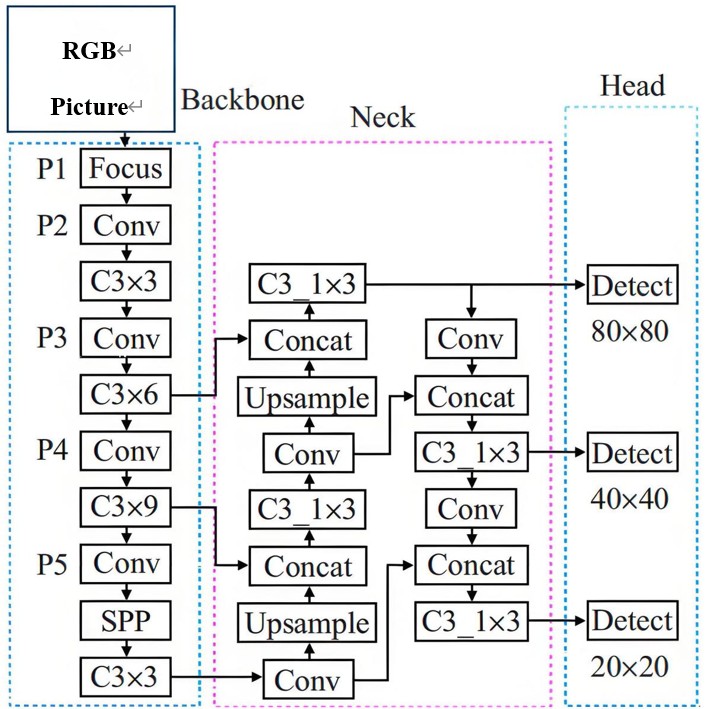}
	  \caption{Model Structure of YOLOv5}\label{FIG:4}
\end{figure}

The total number of images used in this project is 10729. The dataset is divided into training, validation and test in 70:15:15 manner, with a total of 7682 images in the training, 1470 images in the validation, and 1577 images in the test. There are two detection categories, including person and vehicle. To ensure the accuracy of the training results, the dataset used for the training model needs to be different from the dataset used for the pre-training model. In this project, one part of the training dataset used is the VOC dataset, and the other part is the outdoor environment images downloaded from public resources in the web.

In order to enable the model to be deployed on microcontroller devices and show better detection performance, the model needs to be compressed. The main model compression methods used in this project are quantization and pruning, also using distillation to improve model accuracy.
Quantization is the process of reducing the precision of the parameters of a neural network model, usually by replacing floating-point numbers (such as float32) with lower-precision data types (e.g., int8 or float16). By reducing the bit width of the values, the memory and storage requirements of the model can be reduced, while reducing the amount of computation and speeding up the model inference. This is valuable on devices with limited computational resources, such as in mobile devices and embedded systems.

Pruning, also known as network pruning, is mainly used to compress the convolutional and fully connected layers in a neural network model, and gradually evolves to prune complex network structures. The idea is to remove unimportant network structures by evaluating the importance of different network structures in the network model with a certain network structure importance evaluation index, including weights, channels and other components of the network model. According to the different structural scales of network pruning, pruning is categorized into channel pruning and weight pruning, which correspond to the pruning methods with channels and weights as the smallest unit, respectively. Channel pruning pruning fine-grained coarser, its disadvantage is that the adjustment of the network structure is more sloppy, resulting in poorer performance compared to weight pruning; its advantage is that it can be achieved on any hardware platform that supports general matrix operations to accelerate training and reasoning, does not rely on the support of external devices. Weight pruning has a finer pruning granularity, and its advantage is that it can adjust the network structure more finely, resulting in better performance than channel pruning. Its disadvantage is that each convolutional kernel of the network still maintains the initial size after pruning, and only part of the weights are set to 0, and the acceleration of training and reasoning on hardware platforms relies on the support of the hardware gas pedal of sparse computing.\cite{5} Considering the need for this project, channel pruning was chosen.

Distillation is to utilize a large network model with high complexity and good performance as a teacher network to guide the learning of a student network with lower complexity, so that the performance of the student network can be as close as possible to the performance of the teacher network, and to achieve the compression goal of compressing the teacher network into a student network. Knowledge distillation can compress a large teacher network into a small network, which is convenient for deployment in resource-constrained environments such as mobile platforms and embedded devices, and is easy to use in combination with other compression methods to achieve a greater degree of model compression.\cite{6} In this project, distillation technique is mainly used to improve the selected model’s overall performance.

\begin{figure}[h]
	\centering
		\includegraphics[scale=1]{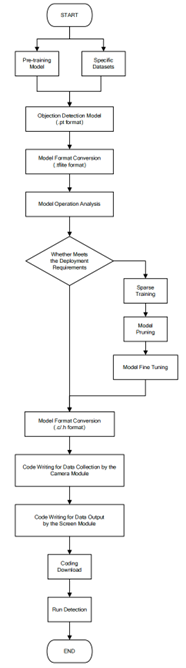}
	  \caption{Experimental Path}\label{FIG:5}
\end{figure}

The implementation of this project can be summarized as Fig.5:
i. Training Model: The target detection model is trained using YOLOv5n pre-trained model, specific dataset, and the format of the trained model is .pt.ii. Model pruning: Using channel pruning to reduce the number of model parameters. Before pruning the model, the model needs to be sparsely trained to identify the less important channels in the model, and then these channels are pruned, in this way reducing the computation while minimizing the impact on the detection accuracy of the model. After pruning, the model needs to be fine-tuned in order to restore its detection capability. If the model still has a large number of parameters, multiple rounds of pruning are required.iii. Distillation: The accuracy of the pruned model will be reduced to a certain extent, and in order to restore the model accuracy as much as possible, distillation is used to improve its detection performance.
iv. Model quantization: the model is quantized to int8 format to simplify the arithmetic process, which subsequently reduces the amount of computation and improves the detection speed. Then converted to tflite format that can be recognized by the STM32MX-CUBE-AI tool, in order to deploy the model running on microcontroller devices.v. Import the model in tflite format into the STM32MX-CUBE-AI software for model analysis, to estimate the size of the model, the space needed for running, etc. After the model passes the evaluation, use STM32MX software to convert it into a detection model and weights file in C language.vi. Write the code for each module function, compile it using Keil5 software and then download the code to the microcontroller.vii. Run the code and perform the detection task.

There are two possible reasons why the AI model cannot pass the evaluation: excessive FLASH usage or excessive RAM usage. If FLASH is too large, it means that the amount of parameters is too large and the number of parameters in the model needs to be reduced. The number of parameters can be reduced by pruning. The analysis of the model evaluation report generated by STM32MX-CUBE-AI reveals that pruning the model mainly reduces the number of parameters, in order to reduce the FLASH space requirements, while there is no significant change in the RAM space requirements. Following this, it was concluded that the factor affecting the RAM space needs is mainly the size of the image to be detected rather than the number of parameters of the model.

When evaluating the performance of the YOLO model, commonly used testing metrics include Precision, Recall, mAP@0.5 and mAP@0.5:0.95. Precision is the ratio of the number of correctly predicted samples to the total number of samples in the classification model, reflecting the overall predictive power of the model. It is calculated as the ratio of the sum of correctly predicted true positive samples (TP) and true negative samples (TN) to the total number of samples.

$$
Precision=\frac{TP+TN}{TP+TN+FP+FN} 
$$

TP (True Positives) The number of correctly predicted positive samples.TN (True Negatives) Number of negative samples correctly predicted.FP (False Positives) The number of negative samples misclassified as positive (false positives).FN (False Negatives) Number of positive samples that is wrong classified as negative (under reporting).Recall is a measure of the proportion of positive samples correctly identified by the model as a proportion of all actual positive samples, reflecting the risk of “missed detection”. It is calculated as the ratio of the number of correctly predicted positive samples to the number of actual positive samples.

$$
Recall=\frac{TP}{TP+FN} 
$$

mAP (mean Average Precision) is a core indicator of the model's detection accuracy, which comprehensively evaluates the model's accuracy by synthesizing the performance under different categories and different detection confidence levels. AP (Average Precision) is the average precision for a single category, which is obtained by calculating the area of the Precision-Recall (PR) curve of the category under different IoU thresholds. mAP is the result of averaging the AP values of all the categories in the dataset to reflect the overall performance of the model in multi-category detection.[1]IoU (Intersection over Union) is a measure of the degree of overlap between the predicted box and the real box, calculated as the ratio of the area of the intersection between the predicted box and the real box to the total area of the predicted box and the real box.

$$
IOU=\frac{Predicted Box\cap{Ground Truth Box}}{Predicted  Box\cup{Ground Truth Box}}
$$

mAP@0.5 is the average accuracy calculated at a single IoU threshold of 0.5. This means that the prediction is only considered correct if the ratio of overlapping areas between the predicted and real boxes is greater than or equal to 0.5. mAP@0.5:0.95: This metric is even more stringent, where AP values are calculated one by one over a range of IoU thresholds (from 0.5 to 0.95 in steps of 0.05), i.e., 0.5, 0.55, 0.6, ..., 0.95, and the average of these AP values is taken as the final result.\cite{7}

When training the model, the commonly used image sizes are 640*640, 320*320, 240*240, 224*224, 192*192, etc,. Larger image size is more likely to obtain a higher detection accuracy. However, too large image size could not work on embedded devices. This project uses RAM limited value to find out available input size range. The test results show that when use 240*240 size image for model training, the RAM usage is about 576.12KB, which makes it difficult to cache the images captured by the camera anymore. Therefore, in order to balance the performance of the hardware device and the detection accuracy, two image sizes are selected for testing in this project as input size, 224*224 and 192*192.
After test, the RAM usage is about 392KB when using the 224*224 image size for detection, and about 288KB when using the 192*192 image size. Both two models needs the same FLASH storage space, with 1.85M.In addition, this project also compares the memory occupation of the model with 192*192 size after pruning, and from the evaluation results, the FLASH requirements space is reduced from 1.85M to 1.75M, but the RAM occupation space is only reduced by 44B.

\begin{figure}[h]
	\centering
		\includegraphics[scale=0.6]{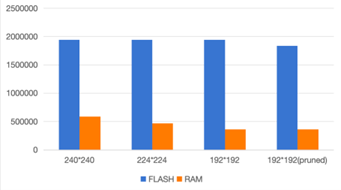}
	  \caption{Hardware storage requirements for different input image sizes}\label{FIG:6}
\end{figure}

The factors affecting the model’s performance (using precision, recall, mAP@0.5, mAP@[0.5:0.95] value to assess) are mainly the input image size and the model’s structure. To realize how these two factors will impact model’s performance and find out a model with overall better performance, this project uses a control variable approach. The independent variables are input image size, model structure, and the dependent variable is th overall detection performance, assessed by the value of mAP@0.5 and mAP@0.5:0.95. The running speed is also a part of assessment.Control the model structure unchanged and test the model accuracy under different input image sizes:In this part ,all the models will not be pruned. The test result shows that input image size with 224*224 obtain a higher overall performance. The value of mAP@0.5 on test datsaset is about 2\% higher than the model with 192 input size, and the value of mAP@0.5:0.95 on test dataset is around 3\%.

\begin{table*}[h]
\caption{Performance (Model with 224*224 input size)}\label{tbl1}
\begin{tabular*}{\tblwidth}{llllll}
\toprule
 TYPE & Number of picture & Precision& Recall &mAP@0.5 &mAP@0.5:0.95  \\ 
\midrule
 All&1577&0.82&0.689& 0.776&0.523  \\
 Vehicle&1577&0.791&0.7025& 0.7735&0.5425  \\ 
 Person&1577&0.789&0.695&0.764&0.477  \\
\bottomrule
\end{tabular*}
\end{table*}

\begin{table*}[h]
\caption{Performance (Model with 192*192 input size))}\label{tbl2}
\begin{tabular*}{\tblwidth}{llllll}
\toprule
 TYPE & Number of picture & Precision& Recall &mAP@0.5 &mAP@0.5:0.95  \\ 
\midrule
 All&1577&0.811&0.681& 0.758&0.497  \\
 Vehicle&1577&0.836&0.6855&0.7825&0.5455  \\ 
 Person&1577&0.772&0.693&0.742&0.446  \\
\bottomrule
\end{tabular*}
\end{table*}

\begin{figure}[h]
	\centering
		\includegraphics[scale=0.6]{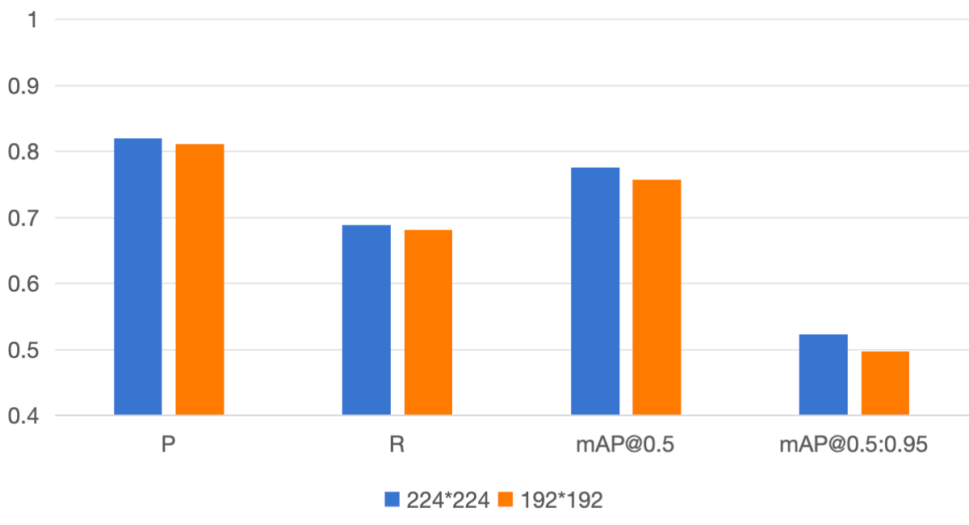}
	  \caption{Performance of the Models (On Test Dataset)}\label{FIG:7}
\end{figure}

From the detection results, it can be seen that the larger the input image size the higher overall performance.(2) Control the model output image size unchanged, reduce the number of model parameters to test the model performance:In order to reduce the number of model parameters, structured pruning is performed on the model with 192*192 input size, and test whether the running speed can be significantly reduced after pruning the model. The original model structure is not suitable for large-scale pruning. In order to minimize the impact on the model detection accuracy, this project only prunes the convolutional modules in the backbone part of the model structure, with a pruning ratio of 10\%. After pruning, the number of channels is retained at 90\%. Before pruning, the model is sparse trained to reduce the model's dependence on unimportant channels. To improve the detection accuracy, the knowledge distillation technique was chosen to optimize the model, using YOLOv5l model as the teacher model in distillation. The performance of each model on test dataset shows as Table3-4:

\begin{table*}[h]
\caption{ Performance (Model with 192*192 input size, pruned)}\label{tbl3}
\begin{tabular*}{\tblwidth}{llllll}
\toprule
 TYPE & Number of picture & Precision& Recall &mAP@0.5 &mAP@0.5:0.95  \\ 
\midrule
 All&1577&0.761&0.547& 0.634&0.376  \\
 Vehicle&1577&0.7825	&0.5495& 0.6435&0.412  \\ 
 Person&1577&0.72&0.544&0.614&0.302  \\
\bottomrule
\end{tabular*}
\end{table*}

\begin{table*}[h]
\caption{Performance (Model with 192*192 input size, pruned and distilled)}\label{tbl4}
\begin{tabular*}{\tblwidth}{llllll}
\toprule
 TYPE & Number of picture & Precision& Recall &mAP@0.5 &mAP@0.5:0.95  \\ 
\midrule
 All&1577&0.735	&0.553&0.631&0.38  \\
 Vehicle&1577&0.7485	&0.557&0.64&0.413  \\ 
 Person&1577&0.708 &0.544&0.613&0.314 \\
\bottomrule
\end{tabular*}
\end{table*}

\begin{figure}[h]
	\centering
		\includegraphics[scale=0.6]{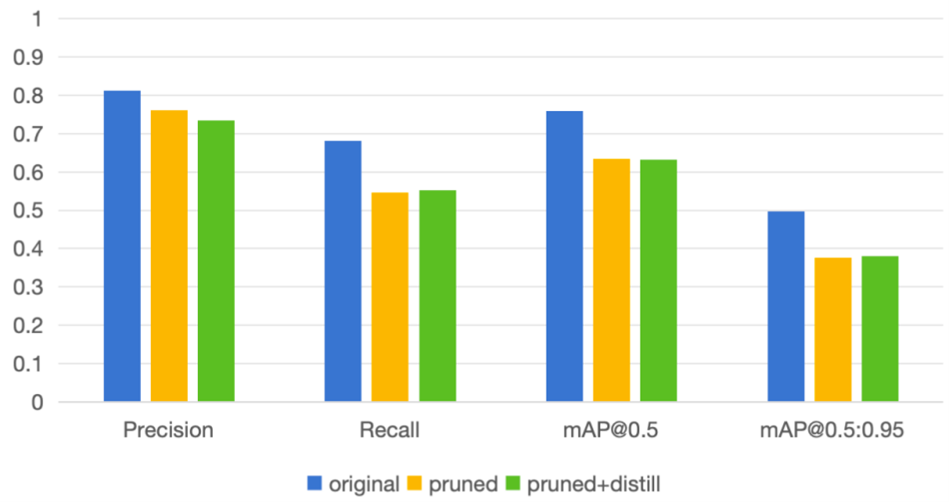}
	  \caption{Performance of the Models (On Test Dataset)}\label{FIG:8}
\end{figure}

\begin{table*}[h]
\caption{Performance (Model with 192*192 input size, distilled without pruned)}\label{tbl5}
\begin{tabular*}{\tblwidth}{llllll}
\toprule
 TYPE & Number of picture & Precision& Recall &mAP@0.5 &mAP@0.5:0.95  \\ 
\midrule
 All&1577&0.812&0.693&0.771&0.519  \\
 Vehicle&1577&0.8285	&0.6875&0.779&0.5465 \\ 
 Person&1577&0.778 &0.703&0.754&0.465 \\
\bottomrule
\end{tabular*}
\end{table*}

\begin{figure}[h]
	\centering
		\includegraphics[scale=0.4]{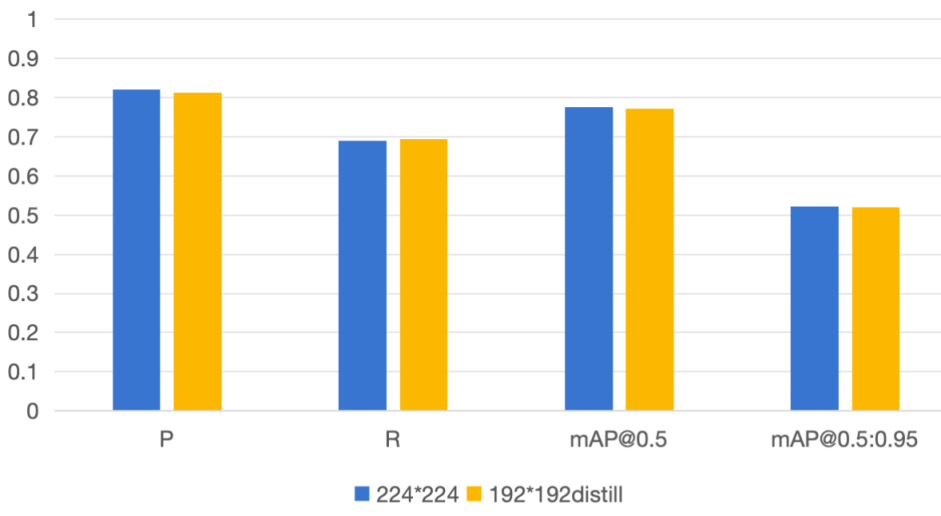}
	  \caption{Performance of the Models (On Test Dataset)}\label{FIG:9}
\end{figure}

Compare to the original model, using distillation still can’t improve the performance of pruned model significantly. The overall performance of model with distillation is nearly the model without distillation. This means even a small ration of pruning still have a severe negative impact on model’s performance. The mAP@0.5 value is 12\% lower than the orginal model and the mAP@0.5:0.9 value is 13\% lower than the original one.
However, when performed distillation process on no pruned model with 192*192 input size, the overall performance is nearly the same as the model with 224*224 input size. The model’s performance on test dataset shows as Table 5:

The model with 192*192 and distilled obtain the mAP@0.5 value of 77.1\%, 0.05\% less than the model with 224*224 input size, and mAP@0.5:0.9 value is 0.04\% less than the model with 224 input size. Distillation improve the model’s performance about 2\%.After model performance evaluation, three model are selected to test their operation speed:(1) Model with input image size of 224*224.(2) Model with input image size of 192*192 after distillation processing.(3) Model with input image size of 192*192 after pruning and distillation processing.The detection method is to start timing after acquiring a image frame, and run the detection code, timing again after completing the detection task. Each model selects 10 valid detections and records the duration, then takes the average value as the running duration of this model. The results are as Fig.10. shows.Model with 224*224 image input size needs to run about 651ms per frame, while two models with 192*192 image input size requires around 490ms. Compare to the whole-structured model with 192*192 image input size, pruned model doesn’t show a significant shorter time requirements.

\begin{figure}[h]
	\centering
		\includegraphics[scale=0.6]{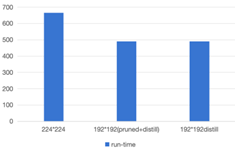}
	  \caption{Running time of models}\label{FIG:10}
\end{figure}

\section{RESULTS}

From the test results, it can be seen that the runtime of the model with input image size of 224*224 is significantly higher than the model with input image size of 192*192 size. Among the models with the same input image size, the pruned model did not significantly improve the running speed, while the model accuracy decreased. After comprehensively evaluating the above information, the model with input image size of 192*192 after distillation processing is selected as the final deployment model. The performance (P-R curve) of selected model on validation dataset and test dataset are show as Fig.11 and Fig 12.Based on the above study, the final model deployed on the hardware device is the model with the input image size of 192*192 after distillation processing, and the overall precision of the model before quantization is 77.4\%, the recall rate is 68.9\%, the value of mAP@0.5 is 75.3\% and the value of mAP@[0.5:0.95] is 50.8\%.

\begin{figure}[h]
	\centering
		\includegraphics[scale=1]{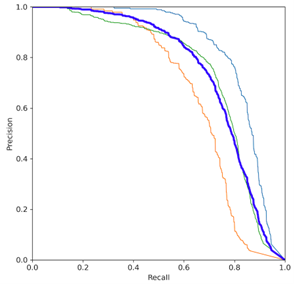}
	  \caption{P-R Curve of model (on validation dataset)}\label{FIG:11}
\end{figure}

\begin{figure}[h]
	\centering
		\includegraphics[scale=1]{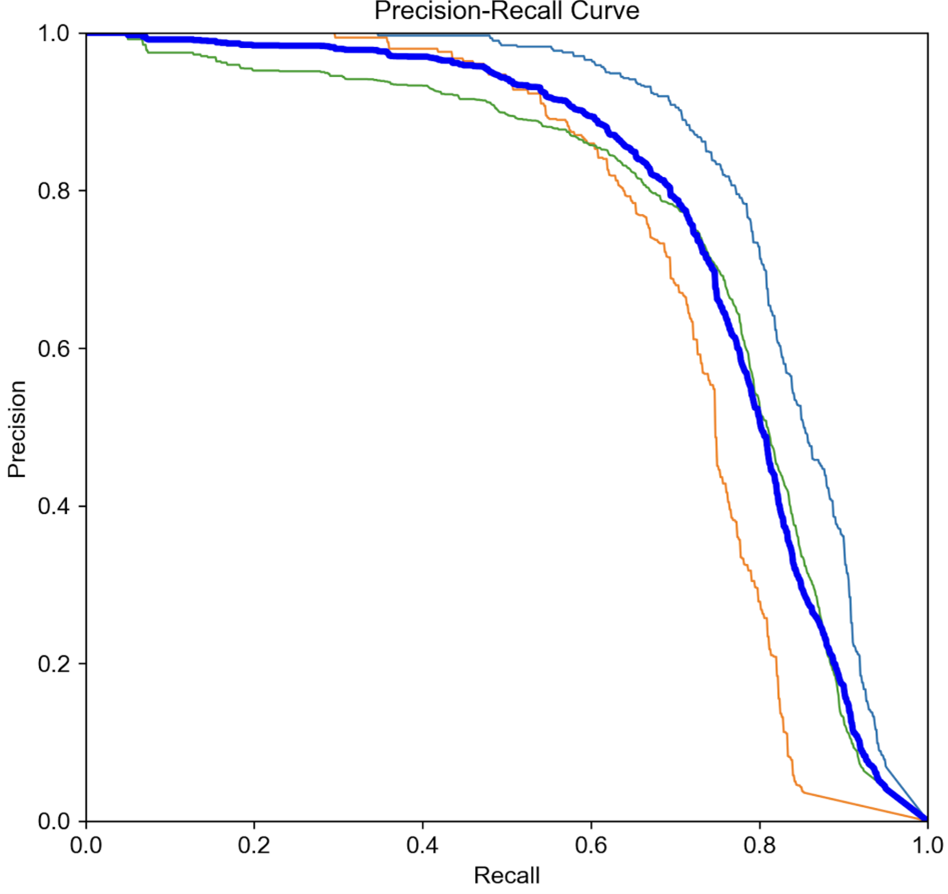}
	  \caption{ P-R Curve of model (on test dataset)}\label{FIG:12}
\end{figure}

\begin{figure*}[h]
	\centering
		\includegraphics[scale=1]{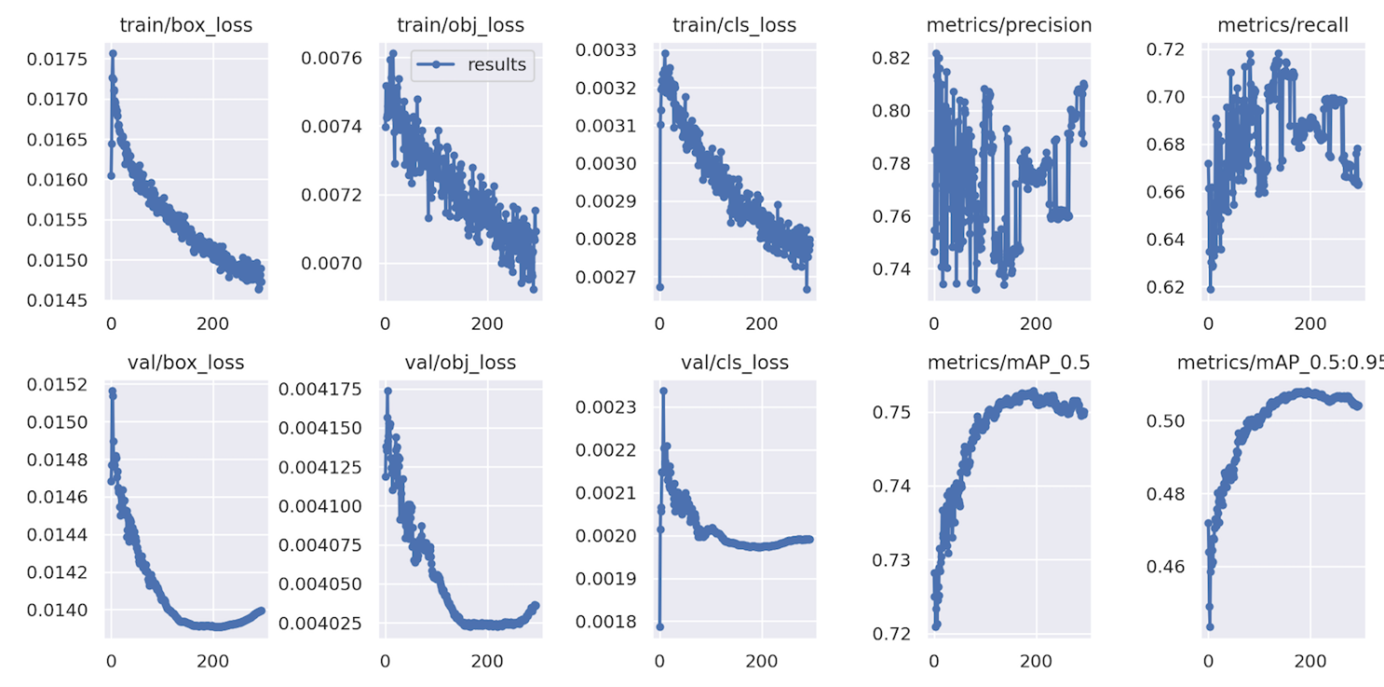}
	  \caption{Records of training process}\label{FIG:13}
\end{figure*}

\section{Conclusion}

This project aims to realize the AI detection model running on low-power devices to achieve portable against the current AI model running equipment requirements are high, energy consumption, the detection category includes people and vehicles. This project starts from the selection of hardware devices, base on the performance, selects STM32H7, which has better performance in low-power MCUs, as the hardware platform. In the selection of the detection model, the model’s detection speed, detection accuracy, and model complexity are considered comprehensively, thus YOLOv5 is finally determined as the basic architecture. In the selection of model parameters, the limitations of the operating conditions of the hardware device are first evaluated, then test two factors that affecting the performance of model, namely the input image size and the model’s structure. The influence of different variables on model performance is analyzed through the control variable method, so as to find out a suitable model that could balance overall performance and running time, in order to  deployment.

After validation, the selected detection model is able to successfully run on the STM32H7 device and perform two categories of detection. The whole set of devices can be powered by mobile power devices (e.g., rechargeable batteries), and suitable for portable devices or for use in outdoor environments without sufficient electric energy (such as in the field).










\bibliographystyle{cas-model2-names}

\bibliography{cas-refs}



\end{document}